\documentclass[11pt]{article}
\usepackage{graphicx}
\usepackage[margin=1in]{geometry}
\usepackage{caption}
\usepackage{subcaption}
\usepackage{amsmath}
\begin{document}

\begin{titlepage}
\begin{center}
\hfill YITP-14-44\\[2cm]

{ \huge \bfseries Calculating the mass fraction of primordial black holes}\\[1cm]

Sam Young $^{a}$, Christian T. Byrnes $^{a}$, Misao Sasaki $^{b}$\\[0.5cm]
$^{a}$Department of Physics and Astronomy, Pevensey II Building, University of Sussex, BN1 9RH, UK\\
$^{b}$ Yukawa Institute for Theoretical Physics, Kyoto University, Kyoto 606-8502, Japan\\[0.5cm]
E-mail: sy81@sussex.ac.uk, cbt22@sussex.ac.uk, misao@yukawa.kyoto-u.ac.jp\\[1cm]

\today\\[1cm]

\end{center}

We reinspect the calculation for the mass fraction of primordial black holes (PBHs) which are formed from primordial perturbations, finding that performing the calculation using the comoving curvature perturbation $\mathcal{R}_{c}$ in the standard way vastly overestimates the number of PBHs, by many orders of magnitude. This is because PBHs form shortly after horizon entry, meaning modes significantly larger than the PBH are unobservable and should not affect whether a PBH forms or not - this important effect is not taken into account by smoothing the distribution in the standard fashion. We discuss alternative methods and argue that the density contrast, $\Delta$, should be used instead as super-horizon modes are damped by a factor $k^{2}$. We make a comparison between using a Press-Schechter approach and peaks theory, finding that the two are in close agreement in the region of interest. We also investigate the effect of varying the spectral index, and the running of the spectral index, on the abundance of primordial black holes.

\end{titlepage}

\tableofcontents

\section{Introduction}

It is believed that primordial black holes (PBHs) could have formed in the early universe from the collapse of large density fluctuations, and if so, could have observational implications - either from their gravitational effects, or the effects of their Hawking radiation (see \cite{Carr:2009jm,Josan:2009qn} for recent lists of the constraints). They have not been observed, but this fact is enough that they can be used to constrain the early universe (i.e. \cite{Bugaev:2012ai,Young:2013oia,Green:1997sz,Byrnes:2012yx,Shandera:2012ke}) - and provide the only known tool for probing the primordial universe on extremely small scales (i.e. \cite{Scott:2012kx}). However, the constraints from PBHs on small scales are much weaker than those on cosmological scales, for example, the constraints from the cosmic microwave background from Planck.

During inflation, the Hubble horizon shrinks on a comoving scale, and quantum fluctuations become classical density perturbations once they exit the horizon. Once inflation ends, the horizon begins to grow and perturbations begin to reenter the horizon. If a perturbation is large enough, it will collapse to form a PBH almost immediately after horizon reentry - and there has been extensive research into the nature of this collapse and how large a perturbation must be in order to collapse \cite{Shibata:1999zs,Hawke:2002rf,Niemeyer:1999ak,Musco:2004ak,Musco:2008hv,Musco:2012au,Niemeyer:1997mt}.

Calculations for the critical value of the density contrast, $\Delta$, or comoving curvature perturbation, $\mathcal{R}_{c}$, above which a region will collapse to form a PBH are typically of order 0.5 or 1 respectively - and so an insignificant number of PBHs will form unless the power spectrum on small scales is much larger than on large scales, by several orders of magnitude. This is possible in several models, such as the running mass model \cite{Drees:2011hb}, axion inflation \cite{Bugaev:2013fya}, a waterfall transition during hybrid inflation \cite{Bugaev:2011wy}, from passive density fluctuations \cite{Lin:2012gs}, or during inflation with small field excursions \cite{Hotchkiss:2011gz}. For a recent summary of PBH forming models see \cite{Green:2014faa}. Alternatively, the constraint on the formation criteria can be relaxed during a phase transition in the early universe, causing PBHs to form preferentially at that mass scale (i.e. \cite{Jedamzik:1999am}).

In this paper, we will review the calculation of the PBH abundance. The calculation typically computes the fraction of the universe which is above the critical value - in terms of $\Delta$ or $\mathcal{R}_{c}$. This is typically done using the theory of peaks, which calculates the number density of peaks above the critical value, or a Press-Schechter approach, which computes the volume of the universe above the critical value. in order to calculate the abundance of PBHs on different scales, the distribution is convolved with a smoothing function to smooth out modes smaller than the horizon, whilst leaving the horizon and super-horizon modes. When $\mathcal{R}_{c}$ is used to do the calculation in this manner, the super-horizon modes have a large impact on the calculation - we will argue that they should not affect the calculation and that using $\mathcal{R}_{c}$ can be misleading and give errors of many orders of magnitude compared to using $\Delta$. 

In Section 2, we will discuss the formation criteria for PBHs explaining these arguments, and in Section 3, we will briefly review the calculation of the mass of a PBH dependant on the scale it forms at. In Section 4 we discuss the different ways the abundance of PBHs and constraints on the early universe can be calculated for different models. We conclude our arguments in Section 5.

\section{Formation criteria}

The abundance of PBHs is normally stated in terms of $\beta$, the mass fraction of  the Universe contained within PBHs at the time of their formation. Typically, $\beta$ is given as a function of their mass (which, we will see later, is a function of the time at which they form) - so that $\beta$ can be used to describe the mass spectrum of PBHs. In order to determine whether a region of the early universe will collapse to form a PBH, then typically either the density or curvature of that region is compared to a threshold value, which itself is typically calculated from numerical simulations.

Traditionally, the density contrast $\Delta=\frac{\delta\rho-\rho}{\rho}$ had been used to calculate $\beta$. However, following the paper by Shibata and Sasaki in 1999 \cite{Shibata:1999zs} which calculated the threshold value in terms of a metric perturbation $\psi$, and the paper by Green, Liddle, Malik and Sasaki (GLMS) in 2004 \cite{Green:2004wb}, it became more common to use the comoving curvature perturbation $\mathcal{R}_{c}$ (for example, \cite{Bugaev:2012ai,Shandera:2012ke})\footnote{The comoving curvature perturbation $\mathcal{R}_{c}$ is equal to the curvature perturbation on uniform density slices $\zeta$ on super-horizon scales, and because sub-horizon modes are smoothed out, it is common to use $\zeta$ instead of $\mathcal{R}_{c}$.}.

\begin{figure}[t]
\centering
	\includegraphics[width=\linewidth]{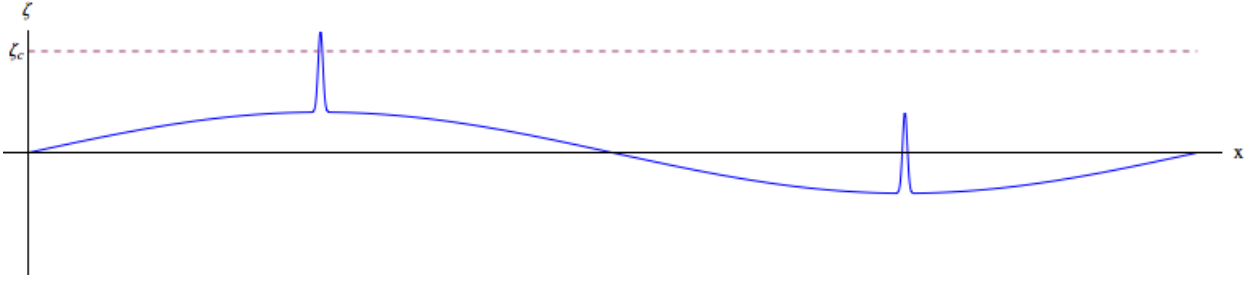}
 \caption{Here, as an example, we show a universe with two sharp (gaussian) peaks in $\mathcal{R}_{c}$ which sit on top of a long wavelength mode. The two thick black boxes represent the size of the visible universe to an observer at the centre of the peaks at the time of PBH formation, whilst the dotted red line represents the hypothetical threshold value for collapse. Both universes appear the same locally to each observer, and so the evolution of each patch should be identical (until the long wavelength becomes observable).}
\label{dontusezeta}
\end{figure}

In figure \ref{dontusezeta} we demonstrate the danger of using $\mathcal{R}_{c}$ to calculate $\beta$. By simply comparing the height of either peak, one would be drawn to the conclusion that the first (left hand) peak will collapse to form a PBH and the second (right hand) peak will not. However, because the long wavelength mode is well outside the horizon, it is unobservable at the expected time of collapse and invoking the separate universe approach (see \cite{Wands:2000dp}) means that it should not affect the local evolution of the universe. Therefore, the universe looks locally identical to observers at either peak - either both peaks should collapse to form a PBH or neither should\footnote{Note that we are assuming that a PBH will form shortly after entering the horizon, or not at all. It is possible for the PBH formation process to last several e-foldings after horizon entry \cite{Musco:2012au} in which case the long wavelength mode will become important, but only for values extremely close to the threshold value - although this is thought to be rare, see equation (\ref{close to critical}) (however, the effect of a perturbation sitting inside a much larger scale perturbation has not been well studied).}.

It should be noted that papers which have calculated a critical value in terms of $\mathcal{R}_{c}$ 
(i.e. \cite{Shibata:1999zs,Nakama:2013ica}) assume that $\mathcal{R}_{c}$ drops quickly to zero outside of the perturbation - so these values can be used if one assumes that there are no super-horizon perturbations affecting your calculation. Therefore it may be possible to use $\mathcal{R}_{c}$ to calculate $\beta$ if one takes care to exclude super-horizon modes from the calculation (one possibility is to simply subtract the long wavelength modes - although this is strongly dependant on what is considered to be a long wavelength.), and  in Section 4.5 we will consider an approximation where only the value of the power spectrum at horizon entry is used.

\begin{figure}[t]
\centering
	\includegraphics[width=\linewidth]{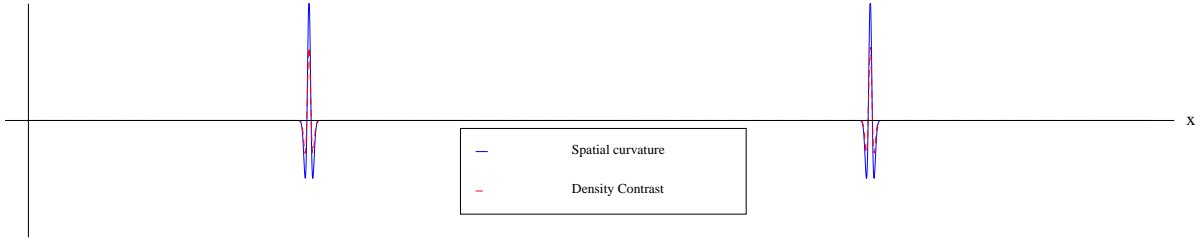}
 \caption{The same universe as shown in figure \ref{dontusezeta}, but this time showing the spatial curvature and the density contrast at the time the scale of the small peaks enter the horizon. We now see that both peaks look identical - and so should evolve in the same manner. We see that the peaks in the spatial curvature and density contrast are very similar, both having a Mexican hat profile (rather than the gaussian shape in the comoving curvature perturbation) - note that the difference in the height of the peaks is due to the arbitrary scaling we have used in the figure.}
\label{curvature density}
\end{figure}

A more formal way to consider this to investigate the effect of super-horizon modes on local observables, such as the density contrast and the spatial 3-curvature. Figure \ref{curvature density} shows the same universe as figure \ref{dontusezeta} but in terms of the spatial curvature and density contrast.

\emph{Spatial curvature} - consider the perturbed, spatially flat FRW metric
\begin{equation}
\label{metric}
ds^{2}=-N^{2}dt^{2}+g_{ij}\left(dx^{i}+N^{i}dt\right)\left(dx^{j}+N^{j}dt\right); g_{ij}=e^{2\alpha}\delta_{ij},
\end{equation}
where we have chosen a comoving slicing, and
\begin{equation}
\alpha=\ln a(t) + \mathcal{R}_{c},
\end{equation}
 with $a(t)$ the scale factor of some flat background and $\mathcal{R}_{c}$ the comoving curvature perturbation. A constant value of $\mathcal{R}_{c}$ can be absorbed into the scale factor by defining
\begin{equation}
\bar{a}(t)=a(t)e^{\mathcal{R}_{c}},
\end{equation}
and so a constant $\mathcal{R}_{c}$ corresponds only to a rescaling of the spatial coordinates, as perhaps clear from the form of the metric (\ref{metric}). The spatial curvature is given by
\begin{equation}
R^{(3)}=-\frac{2}{e^{2\alpha}}\delta^{ij}\left(2\alpha,_{ij}+\alpha,_{i}\alpha,_{j}\right),
\end{equation}
and the spatial curvature of the metric is then
\begin{equation}
\label{spatial curvature}
R^{(3)}=-\frac{2}{e^{2\alpha}}\left(2\nabla^{2}\mathcal{R}_{c}+(\vec{\nabla}\mathcal{R}_{c})^{2}\right).
\end{equation}
If we consider a very long wavelength $\mathcal{R}_{c}$ mode, which appears constant on horizon scales, we see that the spatial curvature due to this mode is negligible due to the derivatives in Eq. (\ref{spatial curvature}).

\emph{Density contrast} - on comoving slices, there is a simple relation at linear order between the comoving curvature perturbation and the density contrast \cite{Green:2004wb}
\begin{equation}
\label{density contrast}
\Delta(t,k)=\frac{2(1+\omega)}{5+3\omega}\left(\frac{k}{aH}\right)^{2}\mathcal{R}_{c}(k),
\end{equation}
where $\omega$ is the equation of state $\omega=p/\rho$, which during radiation domination is $\frac{1}{3}$.\footnote{Josan, Green and Malik \cite{Josan:2009qn} derive an alternative formula valid on super- and sub-horizon scales during radiation domination,
\begin{equation}
\Delta(t,k)=-\frac{4}{\sqrt{3}}\left(\frac{k}{aH}\right)j_{1}\left(\frac{k}{\sqrt{3}aH}\right)\mathcal{R}_{c}(k),
\end{equation}
where $j_{1}$ is a spherical Bessel function. However, after smoothing, there is little difference between this and equation (\ref{density contrast}).} In real space this is
\begin{equation}
\Delta(t,x)=\frac{2(1+\omega)}{5+3\omega}\left(\frac{1}{aH}\right)^{2}\nabla^{2}\mathcal{R}_{c}(x).
\end{equation}
Again, we see that this depends on the second derivative of $\mathcal{R}_{c}$ - and so the effect of super-horizon $\mathcal{R}_{c}$ modes is negligible. At linear order, the density contrast is therefore equivalent to the spatial curvature. However, there has been extensive research into the threshold value of $\Delta$ but not for $R^{(3)}$, we therefore advocate the use of the density contrast in order to calculate the mass fraction, $\beta$.

There has been extensive research on the threshold value for the density contrast above which a PBH will form.  Carr \cite{Carr:1975qj} was the first to derive a threshold value for the formation of PBHs, $\Delta_{c}\approx\omega$ where $\omega$ is the equation of state, by calculating the density necessary for gravity to overcome pressure forces. In recent years, numerical simulations of gravitational collapse have been used to investigate the collapse of different shapes of the initial density profile. Niemeyer and Jedamzik \cite{Niemeyer:1999ak} studied initial shapes including gaussian, Mexican hat, and polynomial, finding $\Delta_{c}\approx0.7$. Musco et al \cite{Musco:2004ak,Musco:2008hv,Musco:2012au}\footnote{Musco et al note that the difference in value obtained by Niemeyer and Jedamzik can be explained because they only considered a pure density perturbation imposed at the time of horizon crossing. Later work included only growing modes accounting for the effect of the perturbation in the velocity field.} later studied PBH formation, finding $\Delta_{c}\approx0.45$. More recently, Harada et al \cite{Harada:2013epa} studied a top hat shape, finding an analytic formula $\Delta_{c}=\sin^{2}[\pi\sqrt{\omega}/(1+3\omega)]=0.41$ during radiation domination, and Nakama et al \cite{Nakama:2013ica} studied generalised shapes to determine the crucial parameters in the shape and size of an overdensity. See also \cite{Hawke:2002rf}. \footnote{It was previously thought that there was an upper bound above which density perturbations would form a separate closed universe rather than a PBH, however, this has been shown not to be the case \cite{Kopp:2010sh}. This is relatively unimportant in practice, as the effect of an upper bound is negligible because higher peaks are exponentially suppressed.}.

\section{Primordial black hole mass}

In order to calculate the mass spectrum, or mass function, of PBHs, it is necessary to relate the horizon scale at the time of formation to the mass of PBH formed. We will first review the calculation of the horizon mass carried out by GLMS \cite{Green:2004wb}. The horizon mass is
\begin{equation}
M_{H}=\frac{4\pi}{3}\rho(H^{-1})^{3}.
\end{equation}
In co-moving units, the horizon scale during radiation domination is $R=(aH)^{-1}\propto a$, and expansion at constant entropy gives $\rho\propto g_{*}^{-1/3}a^{-4}$ (where $g_{*}$ is the number of relativistic degrees of freedom, which is expected to be of order 100 in the early universe). This allows the horizon mass at a given reentry scale to be related to the horizon mass at matter radiation equality,
\begin{equation}
M_{H}=\frac{3}{2}M_{H,eq}(k_{eq}R)^{2}\left(\frac{g_{*,eq}}{g_{*}}\right)^{1/3},
\end{equation}
where we use $k_{eq}=0.07\Omega_{m}h^{2}$Mpc$^{-1}$, $g_{*,eq}\approx 3$ and $g_{*}\approx 100$. $M_{H,eq}$ is given by
\begin{equation}
M_{H,eq}=\frac{4\pi}{3}2\rho_{rad,eq}H_{eq}^{-3}=\frac{8\pi}{3}\frac{\rho_{rad,0}}{k_{eq}^{3}a_{eq}},
\end{equation}
where we take $a_{eq}^{-1}=24 000\Omega_{m}h^{2}$ and $\Omega_{rad,0}h^{2}=4.17\times10^{-5}$. Taking $\Omega_{m}h^{2}=0.14$ gives $M_{H,eq}=7\times10^{50}g$ (for this calculation, we have used the same numbers as GLMS \cite{Green:2004wb}).

Now that the horizon mass has been calculated, it remains to determine fraction of the horizon mass which goes into the PBH, $f_{H}$. Several papers (for example, \cite{Hawke:2002rf,Niemeyer:1999ak}) have noted that, when the density is close to the critical value, the mass of PBH formed depends on the size of the over-density, obeying a simple power law,
\begin{equation}
f_{H}=C\left(\Delta-\Delta_{c}\right)^{\gamma},
\end{equation} 
where $C$ and $\gamma$ are constants - although the values calculated depend on the shape of the initial over-density. Chisholm \cite{Chisholm:2006qc} summarises the different measurements, as well as discussing a minimum bound on the PBH mass from entropy constraints. Typical values for these parameters which we will consider here are $C=3$, $\Delta_{c}=0.5$, and $\gamma=0.3$. For these values, the mass of PBH formed is only significantly smaller than than the horizon mass, $M_{PBH}<0.1M_{H}$, for values of $\Delta$ in the range
\begin{equation}
0.5<\Delta<0.500012,
\label{close to critical}
\end{equation}
and so we will assume that PBHs form with a mass approximately equal to the horizon mass for the remainder of this paper. PBHs of significantly larger mass could form in regions where $\Delta$ is substantially larger than 0.5, but the abundance of these regions is exponentially suppressed, and are thus extremely rare.

\section{Primordial black hole abundance}

We will now discuss the calculation of the PBH mass fraction, $\beta$. The density contrast on a comoving slicing, $\Delta$, is smoothed on a given scale $R$, and the fraction of the universe with a density contrast above the critical value is calculated. The smoothed density contrast $\Delta(R,x)$ is calculated by convolving the density contrast with a window function $W(R,x)$:
\begin{equation}
\Delta(R,x)=\int_{-\infty}^{\infty}d^{3}x' W(R,x-x')\Delta(x').
\end{equation}
The variance of $\Delta(R,x)$ is given by
\begin{equation}
\label{density variance}
\langle\Delta^{2}\rangle=\int_{0}^{\infty}\frac{dk}{k}\tilde{W}^{2}(R,k)\mathcal{P}_{\Delta}(k),
\end{equation}
where $\tilde{W}(R,k)$ is the fourier transform of the window function, and $\mathcal{P}_{\Delta}(k)$ is the density power spectrum. Using equation (\ref{density contrast}) this can be related to the comoving curvature perturbation power spectrum as,
\begin{equation}
\langle\Delta^{2}\rangle=\int_{0}^{\infty}\frac{dk}{k}\tilde{W}^{2}(R,k)\frac{4(1+\omega)^{2}}{(5+3\omega)^{2}}\left(kR\right)^{4}\mathcal{P}_{\mathcal{R}_{c}}(k).
\end{equation}

Throughout this paper, we will use a volume-normalised gaussian window function, such that the fourier transform is given by
\begin{equation}
\label{window function}
\tilde{W}(R,k)=\exp\left(-\frac{k^{2}R^{2}}{2}\right).
\end{equation}

In the remaining portion of this section, we discuss the difference between using a peaks theory or Press-Schechter approach, and the predicted mass spectra of PBHs for a scale invariant curvature spectrum, a power law spectrum and for a spectrum with a running of the spectral index.

\subsection{Peaks theory vs Press-Schechter}

The initial mass fraction of the Universe $\beta$, that went into PBHs can be calculated either using a peaks theory approach, or a Press-Schechter approach. A comparison of these two methods was carried out by GLMS \cite{Green:2004wb} who compared the mass spectra calculated using the curvature perturbation, with peaks theory, and the density contrast, using a Press-Schechter approach. In their calculation it was necessary to assume a blue primordial power spectrum, $n_{s}>1$, and they found the two to be in close agreement\footnote{In the appendix, we correct their calculation, finding that calculating $\beta$ in the different methods disagree strongly.}. We will repeat the calculation here for the density contrast only - finding that using peaks theory or a Press-Schechter are not in as close agreement previously found in \cite{Green:2004wb} but still similar, to within a factor of order 10. 

To investigate the difference between the two methods, we will use a variable $\nu=\Delta/\sigma$\footnote{We note here that with peaks theory, the critical value is stated in terms of the peak value of a fluctuation, but in a Press-Schechter approach, it is the average value of the fluctuation. The relationship between the peak value and the average depends on the shape of the fluctuation - but typically, these are expected to differ only by a factor of order unity, with the peak value being higher. The difference in the critical value of the peak value and the average is therefore within the error of the predicted critical value from different sources. We also note the fact that looking for peaks above a certain value in a smoothed distribution is equivalent to looking for patches with an average density above that value - and so the distinction here is only a technical note.}, where $\sigma$ is the square root of the variance $\langle\Delta^{2}\rangle$ given by equation (\ref{density variance}) and is a function of the form of the power spectrum and the smoothing scale (the calculation of $\sigma$ is the same for either method).

In the theory of peaks, the number density of peaks above a height $\nu_{c}$ is given by \cite{Bardeen:1985tr}
\begin{equation}
n_{peaks}(\nu_{c},R)=\frac{1}{(2\pi)^{2}}\left(\frac{\langle k^{2}\rangle(R)}{3}\right)^{\frac{3}{2}}\left(\nu_{c}^{2}-1\right)\exp\left(-\frac{\nu_{c}^{2}}{2}\right),
\label{peaks}
\end{equation}
where $\langle k^{2}\rangle$ is the second moment of the smoothed density power spectrum
\begin{equation}
\langle k^{2}\rangle(R)=\frac{1}{\langle\Delta^{2}\rangle(R)}\int_{0}^{\infty}\frac{dk}{k}k^{2}\tilde{W}^{2}(k,R)\mathcal{P}_{\Delta}(k).
\end{equation}
If we assume a power law spectrum $\mathcal{P}_{\mathcal{R}_{c}}=A_{\mathcal{R}_{c}}(k/k_{0})^{n_{s}-1}$, and a gaussian window function (equation (\ref{window function})), we obtain
\begin{equation}
\langle k^{2}\rangle(R)=\frac{n_{s}+3}{2R^{2}},
\end{equation}
assuming that $n_{s}>-3$. The number density of peaks above the threshold can be related to the density parameter $\Omega_{PBH,peaks}$ (which is equal to the mass fraction $\beta$ for a flat universe) by $\Omega_{PBH,peaks}(\nu_{c})=n_{peaks}(\nu_{c},R)M(R)/\rho$, where $M(R)$ is the mass of PBH associated with the horizon size $R$, $M(R)=(2\pi)^{3/2}\rho R^{3}$. Finally, we have
\begin{equation}
\beta_{peaks}(\nu_{c})=\Omega_{PBH,peaks}(\nu_{c})=\frac{(n_{s}+3)^{3/2}}{6^{3/2}(2\pi)^{1/2}}\nu_{c}^{2}\exp\left(-\frac{\nu_{c}^{2}}{2}\right).
\label{beta peaks}
\end{equation}

By contrast, the Press-Schechter calculation simply integrates the probability distribution function (PDF), 
\begin{equation}
P(\nu)=\frac{1}{\sqrt{2\pi}}\exp\left(-\frac{\nu^{2}}{2}\right), 
\end{equation}
over the range of values that form a PBH:
\begin{equation}
\beta_{PS}(\nu_{c})=2\int_{\nu_{c}}^{\infty}P(\nu)d\nu=2\int_{\nu_{c}}^{\infty}\frac{1}{\sqrt{2\pi}}\exp\left(-\frac{\nu^{2}}{2}\right)d\nu.
\end{equation}
This can be written in terms of the complimentary error function simply as
\begin{equation}
\beta_{PS}(\nu_{c})=\textrm{erfc}\left(\frac{\nu_{c}}{\sqrt{2}}\right),
\end{equation}
and using the asymptotic expansion of erfc$(\nu_{c})$ this can be written as
\begin{equation}
\beta_{PS}(\nu_{c})\approx\sqrt{\frac{2}{\pi}}\frac{1}{\nu_{c}}\exp\left(-\frac{\nu_{c}^{2}}{2}\right).
\end{equation}

\begin{figure}[t]
\centering
	\includegraphics[width=0.7\linewidth]{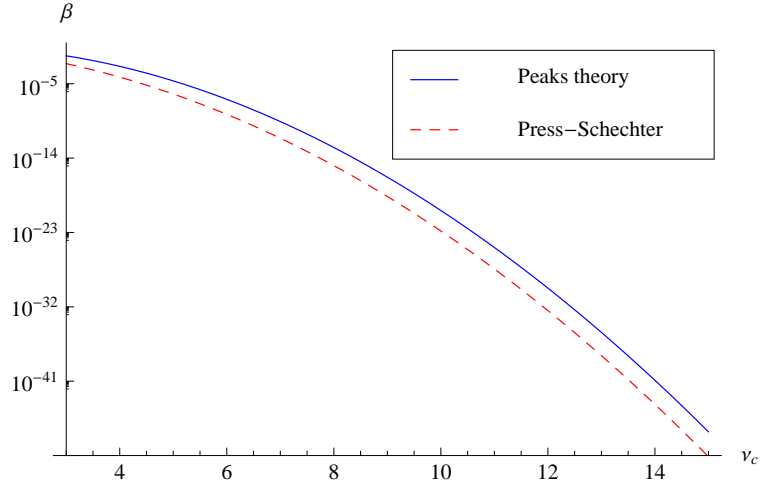}
 \caption{Here we compare the value of $\beta$ calculated using peaks theory or Press-Schechter against $\nu_{c}=\frac{\Delta_{c}}{\sigma}$.}
\label{peaks vs PS}
\end{figure}

Figure \ref{peaks vs PS} shows the difference in the predicted values of $\beta$ for either calculation - the two are in relatively close agreement (differing by a factor of order 10), whilst $\nu$ is not too large\footnote{However, this uncertainty in $\beta$ has little effect on the uncertainty of $\nu_{c}$ which would be calculated, as it depends only on $\log(\beta)$ (see \cite{Young:2013oia}).}. For larger values of $\nu_{c}$, $\beta_{peaks}$ is systematically higher than $\beta_{PS}$. However, the difference between these methods is small compared to the error due to uncertainties in the threshold value $\Delta_{c}$ (see figure \ref{spectral index} for an example).

\subsection{Scale invariant power spectrum}
In the case where the primordial curvature power spectrum is scale invariant, $\mathcal{P}(k)=A_{\mathcal{R}_{c}}$, where $A_{\mathcal{R}_{c}}$ is a constant, then the variance of the smoothed density field during radiation domination, $\omega=1/3$, is
\begin{equation}
\langle\Delta^{2}\rangle=\int_{0}^{\infty}\frac{dk}{k}\tilde{W}^{2}(k,R)\frac{4(1+\omega)^{2}}{(5+3\omega)^{2}}A_{\mathcal{R}_{c}}
=\frac{8}{81}A_{\mathcal{R}_{c}}.
\label{scale invariant density variance}
\end{equation}
Note that, as expected for a scale invariant spectrum, this is now independent of the smoothing scale $R$ 
 - and so predicts that $\beta$ is independent of the mass of the PBHs\footnote{It is also worth noting that for either a red or scale invariant power spectrum $\langle\mathcal{R}_{c}^{2}\rangle\rightarrow\infty$.}. Using peaks theory:
\begin{equation}
\beta=\frac{1}{2^{3/2}(2\pi)^{1/2}}\frac{81\Delta_{c}^{2}}{8A_{\mathcal{R}_{c}}}\exp\left(-\frac{81\Delta_{c}^{2}}{16A_{\mathcal{R}_{c}}}\right).
\label{scale invariant beta} 
\end{equation}

\subsubsection{Constraints on the power spectrum}

Using the relation between the (scale invariant) comoving curvature perturbation power spectrum and $\beta$, equation (\ref{scale invariant beta}), it is simple to calculate a constraint on the power spectrum from the constraint on $\beta$ at a given scale. We will here consider a constraint of size $\beta<10^{-20}$, with $\Delta_{c}=0.5$, and give the constraints one would calculated from peaks theory and Press-Schechter, seeing that the two are in very close agreement:
\begin{align}
\mathcal{P}_{\mathcal{R}_{c},peaks}<0.026, \nonumber \\
\mathcal{P}_{\mathcal{R}_{c},PS}<0.029.
\label{constraints}
\end{align}

\subsection{Power law power spectrum}

\begin{figure}[t]
\centering
	\includegraphics[width=\linewidth]{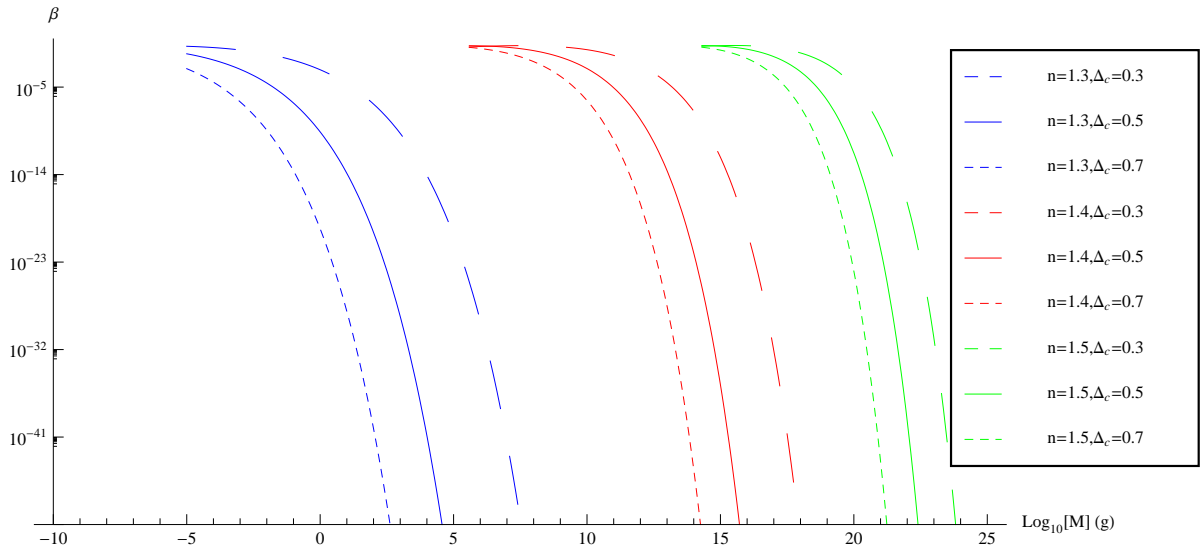}
 \caption{This figure shows the predicted PBH mass spectra for different values of $n_{s}$ and $\Delta_{c}$. A smaller spectral index produces PBHs of smaller masses. Note that the calculation has been artificially cut off when $\beta$ becomes large as it is only valid for rare peaks (where $\beta$ is small), as well as for PBHs smaller than the Planck mass ($M\approx 10^{-5}g$).}
\label{spectral index}
\end{figure}

In order to compare with the GLMS paper \cite{Green:2004wb}, we will consider a power law spectrum (see also Drees and Erfani \cite{Drees:2011hb}). The form of the power spectrum is given by
\begin{equation}
\mathcal{P}_{\mathcal{R}_{c}}(k)=A_{0}\left(\frac{k}{k_{0}}\right)^{n_{s}-1},
\end{equation}
where $A_{0}$ is the amplitude of the power spectrum defined on some pivot scale $k_{0}$, and we will consider only blue spectra, $n_{s}>1$. In this case, the variance of the smoothed density field during radiation domination, given by equation (\ref{density variance}) is
\begin{equation}
\langle\Delta^{2}\rangle=\frac{8}{81}\frac{A_{0}}{(k_{0}R)^{n_{s}-1}}\Gamma\left(\frac{n_{s}+3}{2}\right),
\end{equation}
and $\beta$ is given by equation (\ref{beta peaks}). For the purposes of making a specific calculation we will take $A_{0}=2.2\times10^{-9}$ and $k_{0}=0.05$ Mpc$^{-1}$, loosely based on observations. Figure \ref{spectral index} shows the predicted mass spectra for a range of different spectral indexes $n_{s}$, and threshold values of the density contrast $\Delta_{c}$ - here, we only consider a blue spectrum (it is possible to consider a red spectrum on small scales in which case $\beta$ is larger for more massive PBHs, but a complicated model is needed to produce a significant number of PBHs and be consistent with observations). 

We can place a limit on the spectral index from the observational constraints on the abundance of PBHs - as has been done previously (for example, \cite{Green:1997sz}). Taking $\Delta_{c}=0.5$ and using the constraint $\beta<10^{-20}$ for PBHs in the mass range $10^{8}$g$<M_{PBH}<10^{10}$g \cite{Josan:2009qn}, the constraint on the spectral index is $n_{s}<1.34$. Because there is a minimum mass of PBHs, at the Planck mass, then we can also place a minimum value on $n_{s}$ which is required to form a significant number of PBHs. Approximately 70 efoldings of inflation are required after todays horizon scale exited during inflation in order for the horizon to reach a sufficiently small scale corresponding to the Planck mass. Typical inflationary models predict that the current horizon scale exited the Hubble scale during inflation about 55 efoldings  before the end of inflation \cite{Liddle:2003as}. In that case, the mass contained in the horizon scale at the end of inflation is approximately $e^{30}M_{\rm Planck}\sim 10^8$g. If we require that $\beta>10^{-20}$ for PBHs of mass $M_{PBH}=10^{-5}$g then the spectral index must be $n_{s}>1.26$. In order for a significant number of PBHs to form, then $n_{s}$ must lie in the range
\begin{equation}
1.26<n_{s}<1.34.
\end{equation}

\subsection{Running of the spectral index}

Over the large range of scales considered here, the spectral index is unlikely to be a constant. We will therefore consider a running of the spectral index, $\alpha$, defined as
\begin{equation}
\alpha=\frac{dn_{s}}{d\ln(k)},
\end{equation}
leading to an expression for the comoving curvature perturbation power spectrum given by
\begin{equation}
\mathcal{P}_{\mathcal{R}_{c}}(k)=A_{0}\left(\frac{k}{k_{0}}\right)^{n_{0}-1+\frac{1}{2}\alpha\ln(k/k_{0})},
\end{equation}
where $A_{0}$ and $n_{0}$ are the values of the power spectrum and spectral index respectively, defined at a pivot scale $k_{0}$. If values are given for parameters $k_{0}$, $A_{0}$, $n_{0}$ and $\alpha$ then the PBH mass spectra can be calculated as before, calculating the variance of the smoothed density contrast using equation (\ref{density variance}) and finding $\beta$ using equation (\ref{beta peaks}). 

\begin{figure}[t]
\centering
	\includegraphics[width=\linewidth]{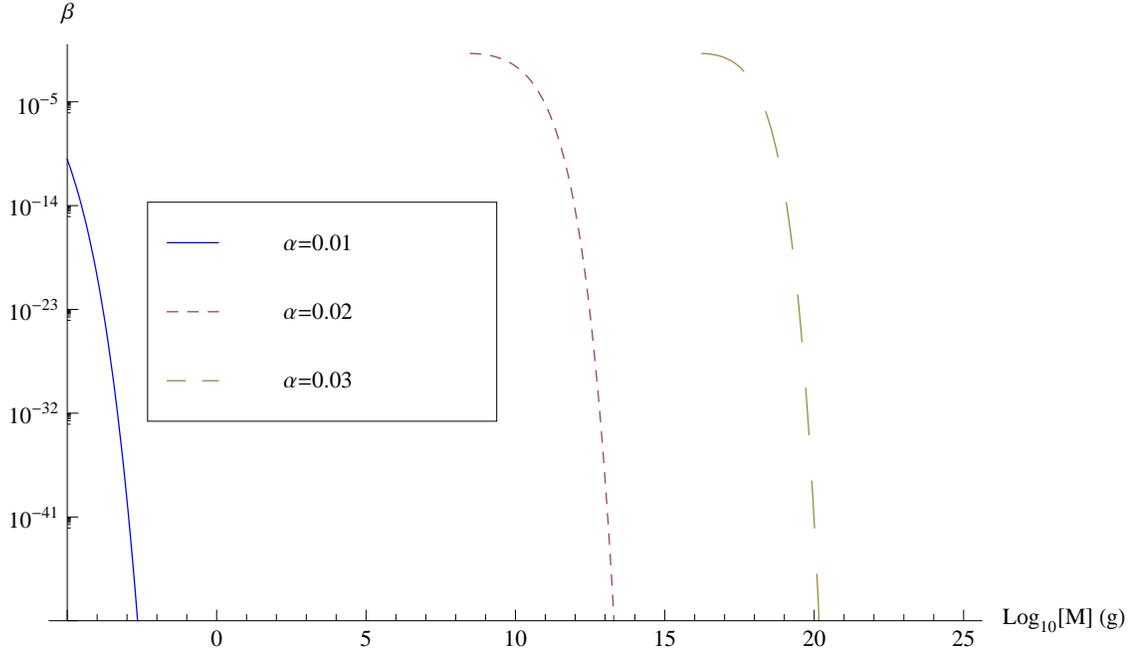}
\caption{This figure shows the predicted PBH mass spectra for different values of the running of the spectral index $\alpha$. Again, the calculation has been artificially cut off when $\beta$ becomes large.}
\label{running}
\end{figure}

The same as in the previous section, we will take $A_{0}=2.2\times10^{-9}$ and $k_{0}=0.05$ Mpc$^{-1}$. The Planck collaboration \cite{Ade:2013uln} found a spectral index $n_{s}=0.9603\pm0.0073$, but no statistically significant running of the spectral index, $\alpha=-0.0134\pm0.0090$. We will therefore take $n_{0}=0.96$ and allow $\alpha$ to vary - see figure \ref{running}. A positive running is necessary to produce a significant number of PBHs, and the smallest value we will consider is $\alpha=0.01$.

PBHs of masses greater than $M_{PBH}\approx 10^{8}$g are well constrained by observations \cite{Josan:2009qn,Carr:2009jm}, and we see from figure \ref{running} that these values of the running produce too many PBHs, and would be ruled out by observational constraints. We therefore state an upper bound on the running of the spectral index, $\alpha<0.0162$ (again, using the constraint $\beta<10^{-20}$ for PBHs in the mass range $10^{8}$g$<M_{PBH}<10^{10}$g \cite{Josan:2009qn}). Although, again, we note that there is no reason to assume the running of the spectral index will be constant over a large range of scales.

We will not consider the running of the running in this paper, although it has been considered by Erfani \cite{Erfani:2013iea}, who places an upper limit on the running of the running by considering the non-production of (long lived) PBHs.

\subsection{Approximation using the comoving curvature perturbation power spectrum}
The power spectrum is, formally, the variance of the amplitude of the Fourier modes at a certain scale. Less formally, one can consider it to be the characteristic size of perturbations at that scale. We show in this section that one can quickly find an approximate value for the PBH mass fraction using the comoving curvature perturbation by only considering perturbations at the exact scale of horizon crossing, without using window functions - this is the approach used in previous papers \cite{Byrnes:2012yx, Young:2013oia}. At horizon crossing, the relation between the density contrast and the comoving curvature perturbation becomes even simpler, as the factor $(k/aH)=1$:
\begin{equation}
\Delta(t_{H},k)=\frac{2(1+\omega)}{5+3\omega}\mathcal{R}_{c}(k)=\frac{4}{9}\mathcal{R}_{c}(k),
\end{equation}
where $t_{H}$ is the time at horizon entry, and $\omega=1/3$ is the equation of state during radiation domination. As $\Delta$ is proportional to $\mathcal{R}_{c}$ at horizon entry, it is reasonable to assume that peaks in the smoothed density contrast correspond to peaks in the comoving curvature perturbation (ignoring other scales).

\begin{figure}[t]
\centering
	\includegraphics[width=\linewidth]{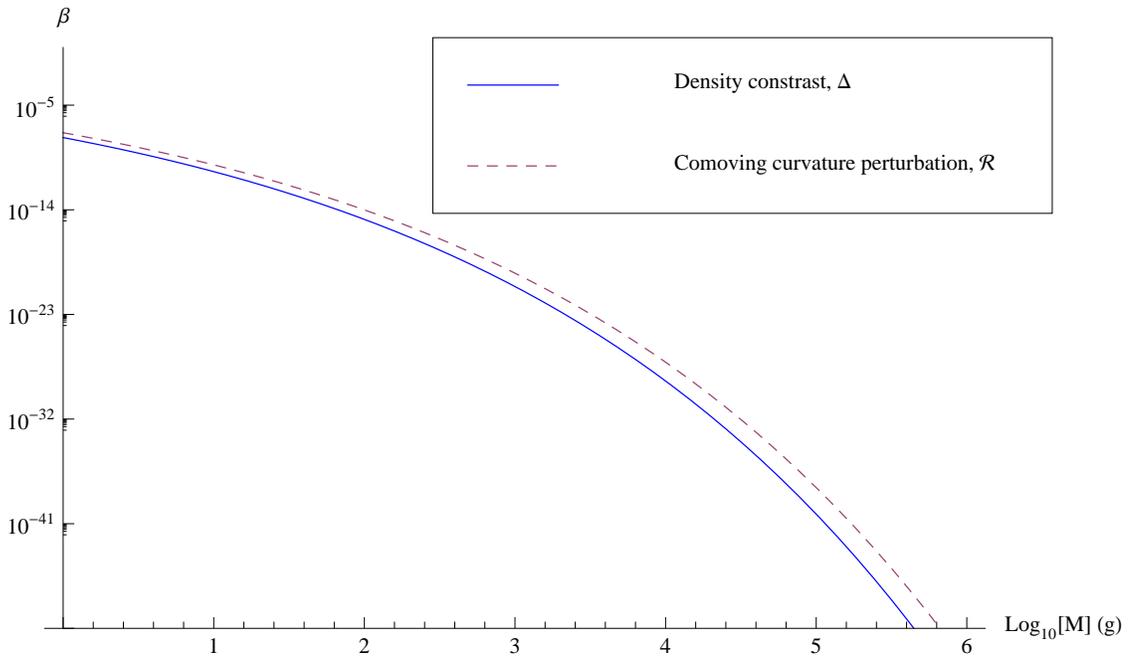}
\caption{We show the mass spectra of PBHs calculated, for a power law power spectrum $\mathcal{P}_{\mathcal{R}_{c}}(k)=A_{0}(k/k_{0})^{n_{s}-1}$, using the density contrast (method described in Section 4.3) and the comoving curvature perturbation (method described in Section 4.5). The values we have used in this figure are $A_{0}=2.2\times10^{-9}$, $k_{0}=0.05$Mpc$^{-1}$, $n_{s}=1.3$, $\Delta_{c}=0.4$ and $\mathcal{R}_{c,crit}=1.2$}
\label{zeta approx and density}
\end{figure}

We will assume that the power spectrum at a given scale gives the variance of the comoving curvature perturbation at that scale and use a Press-Schechter approach to calculate $\beta$:
\begin{equation}
\beta=2\int_{\mathcal{R}_{c,crit}}^{\infty}P(\mathcal{R}_{c})d\mathcal{R}_{c},
\end{equation}
where $P(\mathcal{R}_{c})$ is the (gaussian) probability distribution function. Writing this in terms of the complimentary error function gives
\begin{equation}
\beta=\textrm{erfc}\left(\frac{\mathcal{R}_{c,crit}}{\sqrt{2 \mathcal{P}_{\mathcal{R}_{c}}}}\right).
\label{zeta approx}
\end{equation}
Compare this to the expression one would derive using the density contrast for a scale invariant power spectrum, where $\langle\mathcal{R}_{c}^{2}\rangle$ is given by equation (\ref{scale invariant density variance}),
\begin{equation}
\beta=\textrm{erfc}\left(\frac{9\Delta_{c}}{4\sqrt{\mathcal{P}_{\mathcal{R}_{c}}}}\right).
\label{beta spectral index}
\end{equation}
These two expressions will be exactly equal if $\Delta_{c}\approx\frac{2\sqrt{2}}{9}\mathcal{R}_{c,crit}$. However, these methods cannot be considered identical, which is evident if a power law spectrum is considered, $\mathcal{P}_{\mathcal{R}_{c}}(k)=A_{0}(k/k_{0})^{n_{s}-1}$. Equation (\ref{zeta approx}) is unchanged, but equation (\ref{beta spectral index}) becomes
\begin{equation}
\beta=\textrm{erfc}\left(\frac{9\Delta_{c}}{4\sqrt{\mathcal{P}_{\mathcal{R}_{c}}\Gamma{\left(\frac{3+n_{s}}{2}\right)}}}\right).
\end{equation}
However, provided that $\Gamma\left(\frac{3+n_{s}}{2}\right)\approx 1$ (which is satisfied if $n_{s}\approx 1$) and $\Delta_{c}=\frac{2\sqrt{2}}{9}\mathcal{R}_{c,crit}$, these two expressions will be approximately equal. Figure \ref{zeta approx and density} shows a specific example of these calculations, showing that they still agree closely.

We now compare the constraints on the power spectrum calculated in this method to the constraints calculated earlier (equation (\ref{constraints})). Using $\mathcal{R}_{c,crit}=1.2$ \cite{Shibata:1999zs,Green:2004wb}, and $\beta<10^{-20}$ gives the constraint
\begin{equation}
\mathcal{P}_{\mathcal{R}_{c}}<0.024,
\end{equation}
which is in close agreement with the previously calculated bound, equation (\ref{constraints}).

\section{Conclusions}

We have placed the calculation of the PBH abundance on a more solid grounding. Using the comoving curvature perturbation $\mathcal{R}_{c}$ can be misleading and care needs to be taken if one wishes to use $\mathcal{R}_{c}$ to perform this calculation, due to the effect of super-horizon modes. The problem with using $\mathcal{R}_{c}$ is most easily seen when one considers either a red or scale-invariant power spectrum, which causes the variance of $\mathcal{R}_{c}$ to diverge (it is possible to complete the calculation when a blue spectrum is considered but the results differ drastically from using $\Delta$, see Appendix). We therefore advocate the use of the density contrast to perform the calculation, which does not suffer from the same problem due to the $k^{2}$ dependance of super-horizon modes. In addition, calculations and simulations to calculate the critical threshold for collapse most often use $\Delta$. However, it is more convenient to calculate $\mathcal{R}_{c}$ when studying inflationary models, and finding the constraints on the small scale power spectrum from PBHs - an approximation for $\beta$ can be quickly calculated using $\mathcal{R}_{c}$ if the power spectrum, $\mathcal{P}_{\mathcal{R}_{c}}$, is used rather than using the variance, $\langle\mathcal{R}_{c}^{2}\rangle$ (although this can only ever be an approximation as modes of a similar scale can affect the production of PBHs - which this calculation ignores). It is therefore important that calculations using $\Delta$ or $\mathcal{R}_{c}$ give the same results, and we have provided a method for doing so.

We have considered both a Press-Schechter approach and a peaks theory approach, finding that there is a significant discrepancy between the two - however, this is dwarfed by the error due to uncertainty in the critical value of the density contrast above which PBHs are assumed to form, $\Delta_{c}$. In this paper, we use the peaks theory method, which has a better theoretical grounding. The implications of this paper will be explored further in future papers.

\section{Acknowledgements}
SY is supported by an STFC studentship, and would like to thank Yukawa Institute for Theoretical Physics for its hospitality during a month long stay which was supported by the Bilateral International Exchange Program (BIEP). CB was supported by a Royal Society University Research Fellowship. The authors would like to thank Will Watson, Aurel Schneider, David Seery, Shaun Hotchkiss, Anne Green, Andrew Liddle, John Miller and Ilia Musco for useful discussion which brought about the production of this paper.

\appendix
\section{Appendix}
For completeness, we include the calculation of the PBH mass fraction $\beta$ using the comoving curvature perturbation, and compare it to the calculation using the density contrast. This was initially done by GLMS \cite{Green:2004wb} who incorrectly calculated the density contrast power spectrum at the time of PBH formation - we will now correct the calculation. Assuming a blue power spectrum, $\mathcal{P}_{\mathcal{R}_{c}}=A_{0}\left(k/k_{0}\right)^{n_{s}-1}$ where $n_{s}>1$, the variance of the smoothed comoving curvature perturbation is
\begin{equation}
\langle\mathcal{R}_{c}^{2}\rangle(R)=\int^{\infty}_{0}\frac{dk}{k}\tilde{W}^{2}(k,R)\mathcal{P}_{\mathcal{R}_{c}}(k)=\frac{A_{0}}{2(k_{0}R)^{n_{s}-1}}\Gamma\left(\frac{n_{s}-1}{2}\right).
\end{equation}
The second moment of the power spectrum is given by
\begin{equation}
\langle k^{2}\rangle=\frac{1}{\langle\mathcal{R}_{c}^{2}\rangle(R)}\int^{\infty}_{0}\frac{dk}{k}k^{2}\tilde{W}^{2}(k,R)\mathcal{P}_{\mathcal{R}_{c}}(k)=\frac{n_{s}-1}{2R^{2}},
\end{equation}
leading us to the final expression for $\beta$ using equation (\ref{peaks}) for comoving curvature perturbation instead of density contrast:
\begin{equation}
\beta(R)=\frac{(n_{s}-1)^{3/2}}{6^{3/2}(2\pi)^{1/2}}\frac{\mathcal{R}_{c,crit}^{2}}{\langle\mathcal{R}_{c}^{2}\rangle(R)}\exp\left(\frac{\mathcal{R}_{c,crit}^{2}}{2\langle\mathcal{R}_{c}^{2}\rangle(R)}\right)
\end{equation}

\begin{figure}[t]
\centering
	\includegraphics[width=0.7\linewidth]{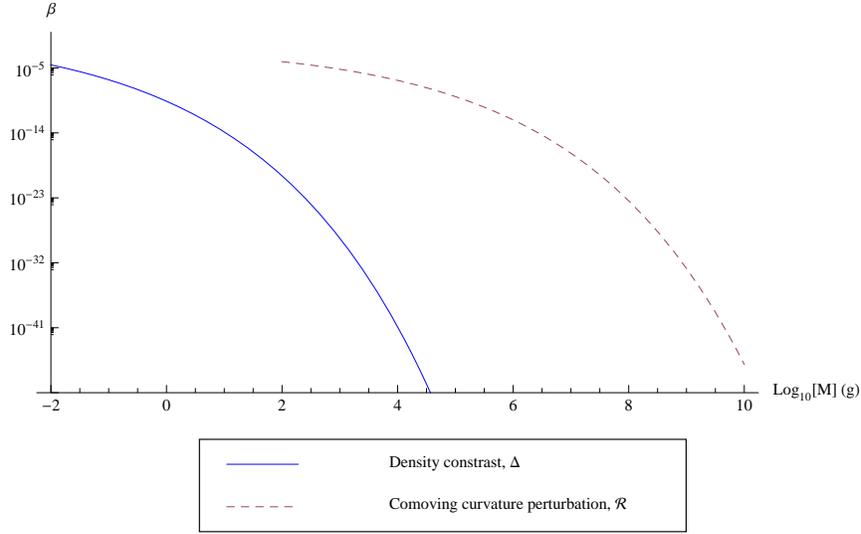}
\caption{We have used $\Delta_{c}=0.5$, $\mathcal{R}_{c,crit}=1$, $n_{s}=1.3$, $A_{0}=2.2\times10^{-9}$ and $k_{0}=0.05$ Mpc$^{-1}$. Both curves represent the mass spectrum of PBHs formed from identical comoving curvature perturbation power spectra - but differ drastically due to the different methods used in the calculation.}
\label{density v curvature}
\end{figure}

The differences between this calculation and the calculation for the density contrast are shown in figure \ref{density v curvature} - we can see that they differ by many orders of magnitude.

\bibliographystyle{JHEP}
\bibliography{bibfile}

\providecommand{\href}[2]{#2}\begingroup\raggedright\begin{thebibliography}{10}

\bibitem{Carr:2009jm}
B.~Carr, K.~Kohri, Y.~Sendouda, and J.~Yokoyama, {\it {New cosmological
  constraints on primordial black holes}},  {\em Phys.Rev.} {\bf D81} (2010)
  104019, [\href{http://xxx.lanl.gov/abs/0912.5297}{{\tt arXiv:0912.5297}}].

\bibitem{Josan:2009qn}
A.~S. Josan, A.~M. Green, and K.~A. Malik, {\it {Generalised constraints on the
  curvature perturbation from primordial black holes}},  {\em Phys.Rev.} {\bf
  D79} (2009) 103520, [\href{http://xxx.lanl.gov/abs/0903.3184}{{\tt
  arXiv:0903.3184}}].

\bibitem{Bugaev:2012ai}
E.~Bugaev and P.~Klimai, {\it {Cosmological constraints on the curvaton web
  parameters}},  {\em Phys.Rev.} {\bf D88} (2013), no.~2 023521,
  [\href{http://xxx.lanl.gov/abs/1212.6529}{{\tt arXiv:1212.6529}}].

\bibitem{Young:2013oia}
S.~Young and C.~T. Byrnes, {\it {Primordial black holes in non-Gaussian
  regimes}},  {\em JCAP} {\bf 1308} (2013) 052,
  [\href{http://xxx.lanl.gov/abs/1307.4995}{{\tt arXiv:1307.4995}}].

\bibitem{Green:1997sz}
A.~M. Green and A.~R. Liddle, {\it {Constraints on the density perturbation
  spectrum from primordial black holes}},  {\em Phys.Rev.} {\bf D56} (1997)
  6166--6174, [\href{http://xxx.lanl.gov/abs/astro-ph/9704251}{{\tt
  astro-ph/9704251}}].

\bibitem{Byrnes:2012yx}
C.~T. Byrnes, E.~J. Copeland, and A.~M. Green, {\it {Primordial black holes as
  a tool for constraining non-Gaussianity}},  {\em Phys.Rev.} {\bf D86} (2012)
  043512, [\href{http://xxx.lanl.gov/abs/1206.4188}{{\tt arXiv:1206.4188}}].

\bibitem{Shandera:2012ke}
S.~Shandera, A.~L. Erickcek, P.~Scott, and J.~Y. Galarza, {\it {Number Counts
  and Non-Gaussianity}},  {\em Phys.Rev.} {\bf D88} (2013) 103506,
  [\href{http://xxx.lanl.gov/abs/1211.7361}{{\tt arXiv:1211.7361}}].

\bibitem{Scott:2012kx}
P.~Scott, T.~Bringmann, and Y.~Akrami, {\it {Constraints on small-scale
  cosmological perturbations from gamma-ray searches for dark matter}},  {\em
  J.Phys.Conf.Ser.} {\bf 375} (2012) 032012,
  [\href{http://xxx.lanl.gov/abs/1205.1432}{{\tt arXiv:1205.1432}}].

\bibitem{Shibata:1999zs}
M.~Shibata and M.~Sasaki, {\it {Black hole formation in the Friedmann universe:
  Formulation and computation in numerical relativity}},  {\em Phys.Rev.} {\bf
  D60} (1999) 084002, [\href{http://xxx.lanl.gov/abs/gr-qc/9905064}{{\tt
  gr-qc/9905064}}].

\bibitem{Hawke:2002rf}
I.~Hawke and J.~Stewart, {\it {The dynamics of primordial black hole
  formation}},  {\em Class.Quant.Grav.} {\bf 19} (2002) 3687--3707.

\bibitem{Niemeyer:1999ak}
J.~C. Niemeyer and K.~Jedamzik, {\it {Dynamics of primordial black hole
  formation}},  {\em Phys.Rev.} {\bf D59} (1999) 124013,
  [\href{http://xxx.lanl.gov/abs/astro-ph/9901292}{{\tt astro-ph/9901292}}].

\bibitem{Musco:2004ak}
I.~Musco, J.~C. Miller, and L.~Rezzolla, {\it {Computations of primordial black
  hole formation}},  {\em Class.Quant.Grav.} {\bf 22} (2005) 1405--1424,
  [\href{http://xxx.lanl.gov/abs/gr-qc/0412063}{{\tt gr-qc/0412063}}].

\bibitem{Musco:2008hv}
I.~Musco, J.~C. Miller, and A.~G. Polnarev, {\it {Primordial black hole
  formation in the radiative era: Investigation of the critical nature of the
  collapse}},  {\em Class.Quant.Grav.} {\bf 26} (2009) 235001,
  [\href{http://xxx.lanl.gov/abs/0811.1452}{{\tt arXiv:0811.1452}}].

\bibitem{Musco:2012au}
I.~Musco and J.~C. Miller, {\it {Primordial black hole formation in the early
  universe: critical behaviour and self-similarity}},  {\em Class.Quant.Grav.}
  {\bf 30} (2013) 145009, [\href{http://xxx.lanl.gov/abs/1201.2379}{{\tt
  arXiv:1201.2379}}].

\bibitem{Niemeyer:1997mt}
J.~C. Niemeyer and K.~Jedamzik, {\it {Near-critical gravitational collapse and
  the initial mass function of primordial black holes}},  {\em Phys.Rev.Lett.}
  {\bf 80} (1998) 5481--5484,
  [\href{http://xxx.lanl.gov/abs/astro-ph/9709072}{{\tt astro-ph/9709072}}].

\bibitem{Drees:2011hb}
M.~Drees and E.~Erfani, {\it {Running-Mass Inflation Model and Primordial Black
  Holes}},  {\em JCAP} {\bf 1104} (2011) 005,
  [\href{http://xxx.lanl.gov/abs/1102.2340}{{\tt arXiv:1102.2340}}].

\bibitem{Bugaev:2013fya}
E.~Bugaev and P.~Klimai, {\it {Axion inflation with gauge field production and
  primordial black holes}},  \href{http://xxx.lanl.gov/abs/1312.7435}{{\tt
  arXiv:1312.7435}}.

\bibitem{Bugaev:2011wy}
E.~Bugaev and P.~Klimai, {\it {Formation of primordial black holes from
  non-Gaussian perturbations produced in a waterfall transition}},  {\em
  Phys.Rev.} {\bf D85} (2012) 103504,
  [\href{http://xxx.lanl.gov/abs/1112.5601}{{\tt arXiv:1112.5601}}].

\bibitem{Lin:2012gs}
C.-M. Lin and K.-W. Ng, {\it {Primordial Black Holes from Passive Density
  Fluctuations}},  {\em Phys.Lett.} {\bf B718} (2013) 1181--1185,
  [\href{http://xxx.lanl.gov/abs/1206.1685}{{\tt arXiv:1206.1685}}].

\bibitem{Hotchkiss:2011gz}
S.~Hotchkiss, A.~Mazumdar, and S.~Nadathur, {\it {Observable gravitational
  waves from inflation with small field excursions}},  {\em JCAP} {\bf 1202}
  (2012) 008, [\href{http://xxx.lanl.gov/abs/1110.5389}{{\tt
  arXiv:1110.5389}}].

\bibitem{Green:2014faa}
A.~M. Green, {\it {Primordial Black Holes: sirens of the early Universe}},
  \href{http://xxx.lanl.gov/abs/1403.1198}{{\tt arXiv:1403.1198}}.

\bibitem{Jedamzik:1999am}
K.~Jedamzik and J.~C. Niemeyer, {\it {Primordial black hole formation during
  first order phase transitions}},  {\em Phys.Rev.} {\bf D59} (1999) 124014,
  [\href{http://xxx.lanl.gov/abs/astro-ph/9901293}{{\tt astro-ph/9901293}}].

\bibitem{Green:2004wb}
A.~M. Green, A.~R. Liddle, K.~A. Malik, and M.~Sasaki, {\it {A New calculation
  of the mass fraction of primordial black holes}},  {\em Phys.Rev.} {\bf D70}
  (2004) 041502, [\href{http://xxx.lanl.gov/abs/astro-ph/0403181}{{\tt
  astro-ph/0403181}}].

\bibitem{Wands:2000dp}
D.~Wands, K.~A. Malik, D.~H. Lyth, and A.~R. Liddle, {\it {A New approach to
  the evolution of cosmological perturbations on large scales}},  {\em
  Phys.Rev.} {\bf D62} (2000) 043527,
  [\href{http://xxx.lanl.gov/abs/astro-ph/0003278}{{\tt astro-ph/0003278}}].

\bibitem{Nakama:2013ica}
T.~Nakama, T.~Harada, A.~Polnarev, and J.~Yokoyama, {\it {Identifying the most
  crucial parameters of the initial curvature profile for primordial black hole
  formation}},  \href{http://xxx.lanl.gov/abs/1310.3007}{{\tt
  arXiv:1310.3007}}.

\bibitem{Carr:1975qj}
B.~J. Carr, {\it {The Primordial black hole mass spectrum}},  {\em
  Astrophys.J.} {\bf 201} (1975) 1--19.

\bibitem{Harada:2013epa}
T.~Harada, C.-M. Yoo, and K.~Kohri, {\it {Threshold of primordial black hole
  formation}},  {\em Phys. Rev.} {\bf D88} (2013) 084051,
  [\href{http://xxx.lanl.gov/abs/1309.4201}{{\tt arXiv:1309.4201}}].

\bibitem{Kopp:2010sh}
M.~Kopp, S.~Hofmann, and J.~Weller, {\it {Separate Universes Do Not Constrain
  Primordial Black Hole Formation}},  {\em Phys.Rev.} {\bf D83} (2011) 124025,
  [\href{http://xxx.lanl.gov/abs/1012.4369}{{\tt arXiv:1012.4369}}].

\bibitem{Chisholm:2006qc}
J.~R. Chisholm, {\it {Primordial Black Hole Minimum Mass}},  {\em Phys.Rev.}
  {\bf D74} (2006) 043512,
  [\href{http://xxx.lanl.gov/abs/astro-ph/0604174}{{\tt astro-ph/0604174}}].

\bibitem{Bardeen:1985tr}
J.~M. Bardeen, J.~Bond, N.~Kaiser, and A.~Szalay, {\it {The Statistics of Peaks
  of Gaussian Random Fields}},  {\em Astrophys.J.} {\bf 304} (1986) 15--61.

\bibitem{Liddle:2003as}
A.~R. Liddle and S.~M. Leach, {\it {How long before the end of inflation were
  observable perturbations produced?}},  {\em Phys.Rev.} {\bf D68} (2003)
  103503, [\href{http://xxx.lanl.gov/abs/astro-ph/0305263}{{\tt
  astro-ph/0305263}}].

\bibitem{Ade:2013uln}
{\bf Planck Collaboration} Collaboration, P.~Ade {\em et.~al.}, {\it {Planck
  2013 results. XXII. Constraints on inflation}},
  \href{http://xxx.lanl.gov/abs/1303.5082}{{\tt arXiv:1303.5082}}.

\bibitem{Erfani:2013iea}
E.~Erfani, {\it {Modulated Inflation Models and Primordial Black Holes}},  {\em
  Phys.Rev.} {\bf D89} (2014) 083511,
  [\href{http://xxx.lanl.gov/abs/1311.3090}{{\tt arXiv:1311.3090}}].

\end{thebibliography}\endgroup

\end{document}